\documentclass{elsart}

\begin{document}

\begin{frontmatter}

\title{Large Alphabets and Incompressibility}
\author{Travis Gagie}
\address{Department of Computer Science\\
        University of Toronto}
\maketitle

\begin{abstract}
We briefly survey some concepts related to empirical entropy --- normal
numbers, de Bruijn sequences and Markov processes --- and investigate how
well it approximates Kolmogorov complexity.  Our results suggest
$\ell$th-order empirical entropy stops being a reasonable complexity
metric for almost all strings of length $m$ over alphabets of size $n$
about when $n^\ell$ surpasses $m$.
\end{abstract}


\end{frontmatter}


\section{Introduction}
\label{introduction}

For data compression, machine learning and cryptanalysis, we often want to
know the \emph{Kolmogorov complexity} \(K
(S)\)~\cite{Sol64,Kol65,Cha69,LV97} of a string $S$, that is, the minimum
space needed to store $S$.  It is formally defined as the length in bits
of the shortest program that outputs $S$.  Notice our choice of
programming language does not affect this length by more than an additive
constant, provided it is Turing-equivalent; for example, the length of the
shortest such {\tt FORTRAN} program exceeds the length of the shortest
such {\tt LISP} program by no more than the length of the shortest {\tt
LISP}-interpreter written in {\tt FORTRAN} --- which does not depend on
$S$.  Unfortunately, a simple diagonalization shows Kolmogorov complexity
is incomputable: Given a program $\mathcal{A}$ for computing Kolmogorov
complexity, we could write a program $\mathcal{B}$ that searches until it
finds and outputs a string $S$ with \(\mathcal{A} (S) = K (S)\) greater
than $\mathcal{B}$'s length in bits, contradicting the definition of \(K
(S)\).  Thus, researchers substitute various other complexity metrics; in
this paper we study one of the most popular --- empirical entropy.

Empirical entropy is rooted in information theory.  Let $X$ be a random
variable that takes on one of $n$ values according to \(P = p_1, \ldots,
p_n\).  Shannon~\cite{Sha48} proposed that any function \(H (P)\)
measuring our uncertainty about $X$ should have three properties:
\begin{enumerate}
\item ``$H$ should be continuous in the $p_i$.''
\item ``If all the $p_i$ are equal, \(p_i = \frac{1}{n}\), then $H$ should 
be a monotonic increasing function of $n$.''
\item ``If a choice be broken down into two successive choices, the 
original $H$ should be the weighted sum of the individual values of $H$.''
\end{enumerate}
He proved the only function with these properties is \(H (P) = \sum_{i =
1}^n p_i \log (1 / p_i)\), which he called the \emph{entropy} of $P$.  The 
choice of the logarithm's base determines the unit; by convention, $\log$ 
means $\log_2$ and the units are bits.

Let $\ell$ be a non-negative integer and suppose \(S = s_1 \cdots s_m\).  
The \emph{$\ell$th-order empirical entropy} of $S$ (see,
e.g.,~\cite{Man01}) is our expected uncertainty about the random variable
$s_i$ given a context of length $\ell$, as in the following experiment:
$i$ is chosen uniformly at random from \(\{1, \ldots, m\}\); if \(i \leq
\ell\), then we are told $s_i$; otherwise, we are told \(s_{i - \ell}
\cdots s_{i - 1}\).  Specifically,
\[H_\ell (S)
= \left\{ \begin{array}{ll}
\displaystyle \sum_{a \in S} \frac{\#_a (S)}{m} \log \frac{m}{\#_a (S)}
\hspace{3ex} & \mbox{if \(\ell = 0\),}\\
& \\
\displaystyle \frac{1}{m} \sum_{|\alpha| = \ell} |S_\alpha| H_0 (S_\alpha)
\hspace{3ex} & \mbox{if \(\ell \geq 1\).}
\end{array} \right.\]

In this paper, \(a \in S\) means character $a$ occurs in $S$; \(\#_a (S)\)  
is the number of occurrences of $a$ in $S$; and $S_\alpha$ is the string
whose $i$th character is the one immediately following the $i$th
occurrence of string $\alpha$ in $S$ --- the length of $S_\alpha$ is the
number of occurrences of $\alpha$ in $S$, which we denote \(\#_\alpha
(S)\), unless $\alpha$ is a suffix of $S$, in which case it is 1 less.  
We assume \(S_\alpha = S\) when $\alpha$ is empty.  Notice \(0 \leq
H_{\ell + 1} (S) \leq H_\ell (S) \leq \log |\{a\,:\,a \in S\}|\) for
\(\ell \geq 0\).  For example, if $S$ is the string TORONTO, then
\begin{eqnarray*}
H_0 (S)
& = & \frac{1}{7} \log 7 +
	\frac{3}{7} \log \frac{7}{3} +
	\frac{1}{7} \log 7 +
	\frac{2}{7} \log \frac{7}{2}
\approx 1.84\ ,\\
&& \\
H_1 (S)
& = & \frac{1}{7} \left( \rule{0ex}{2ex}
	H_0 (S_\mathrm{N})
	+ 2 H_0 (S_\mathrm{O})
	+ H_0 (S_\mathrm{R})
	+ 2 H_0 (S_\mathrm{T})
	\right)\\
& = & \frac{1}{7} \left( \rule{0ex}{2ex}
	H_0 (\mathrm{T})
	+ 2 H_0 (\mathrm{RN})
	+ H_0 (\mathrm{O})
	+ 2 H_0 (\mathrm{OO})
	\right)\\
& = & 2 / 7 \approx 0.29
\end{eqnarray*}
and all higher-order empirical entropies of $S$ are 0.  This means if
someone chooses a character uniformly at random from TORONTO and asks us
to guess it, then our uncertainty is about \(1.84\) bits.  If they tell us
the preceding character before we guess, then on average our uncertainty
is about \(0.29\) bits; if they tell us the preceding two characters, then 
we are certain of the answer.

Empirical entropy has a surprising connection to number theory.  Let
$(x)_{n, m}$ denote the first $m$ digits of the number $x$ in base \(n
\geq 2\).  Borel~\cite{Bor09} called $x$ \emph{normal in base $n$} if, for
\(\alpha \in \{0, \ldots, n - 1\}^*\), \(\lim_{m \rightarrow \infty}
\frac{\#_\alpha ((x)_{n,m})}{m} = 1 / n^{|\alpha|}\).  For example, the
Champernowne constant~\cite{Cha33} and Copeland-Erd\"{o}s
constant~\cite{CE46}, \(0\,.\,1\,2\,3\,4\,5\,6\,7\,8\,9\,10\,11\,12
\ldots\) and \(0\,.\,2\,3\,5\,7\,11\,13\,17\,19\,23 \ldots\), are normal
in base 10.  Notice $x$ being normal in base $n$ is equivalent to
\(\lim_{m \rightarrow \infty} H_\ell ((x)_{n, m}) = \log n\) for \(\ell
\geq 0\).  Borel called $x$ \emph{absolutely normal} if it is normal in
all bases.  He proved almost all numbers are absolutely normal but
Sierpinski~\cite{Sie17} was the first to find an example, which is still
not known to be computable.  Turing~\cite{Tur??} claimed there exist
computable absolutely normal numbers but this was only verified recently,
by Becher and Figueira~\cite{BF02}.  Such numbers' representations have
finite Kolmogorov complexity yet look random if we consider only empirical
entropy --- regardless of base and order.  Of course, we are sometimes
fooled whatever computable complexity metric we consider.

Now consider de Bruijn sequences~\cite{deB46} from combinatorics.  An
\emph{$n$-ary linear de Bruijn sequence of order $\ell$} is a string over
\(\{0, \ldots, n - 1\}\) containing every possible $\ell$-tuple exactly
once.  For example, the binary linear de Bruijn sequences of order $3$ are
the $16$ $10$-bit substrings of \(00010111000101110\) and its reverse:
\(0001011100, \dots, 1000101110, 0111010001, \ldots, 0011101000\).  By
definition, such strings have length \(n^\ell + \ell - 1\) and
$\ell$th-order empirical entropy $0$ (but \((\ell - 1)\)st-order empirical
entropy \(\frac{(n^\ell - 1) \log n}{n^\ell + \ell - 1}\)).  However,
Rosenfeld~\cite{Ros02} showed there are \((n!)^{n^{\ell - 1}}\) of them.  
It follows that one randomly chosen has expected Kolmogorov complexity in
\(\Theta \left( \log (n!)^{n^{\ell - 1}} \right) = \Theta (n^\ell \log
n)\); whereas Borel's normal numbers can be much less complex than
empirical entropy suggests, de Bruijn sequences can be much more complex.

Empirical entropy also has connections to algorithm design.  For example,
Munro and Spira~\cite{MS76} used $0$th-order empirical entropy to analyze
several sorting algorithms and Sleator and Tarjan~\cite{ST85} used it in
the Static Optimality Theorem: Suppose we perform a sequence of $m$
operations on a splay-tree, with $s_i$ being the target of the $i$th
operation; if \(S = s_1 \cdots s_m\) includes every key in the tree, then
we use \(O ((H_0 (S) + 1) m)\) time.  Of course, most of the algorithms
analyzed in terms of empirical entropy are for data compression.  
Manzini's analysis~\cite{Man01} of the Burrows-Wheeler
Transform~\cite{BW94} is particularly interesting.  He proved an algorithm
based on the Transform stores any string $S$ of length $m$ over an
alphabet of size $n$ in at most about \((8 H_\ell (S) + 1 / 20) m + n^\ell
(2 n \log n + 9)\) bits, for all \(\ell \geq 0\) simultaneously.  
Subsequent research by Ferragina, Manzini, M{\"a}kinen and
Navarro~\cite{FMMN??}, for example, has shown that if \(n^{\ell + 1} \log
m \in o (m \log n)\), then we can store an efficient index for $S$ in
\((H_\ell (S) + o (\log n)) m\) bits.  Notice we cannot lift the
restriction on $n$ and $\ell$ to \(n^{\ell} \in O (m)\): If $S$ is a
randomly chosen $n$-ary linear de Bruijn sequence of order $\ell$, then
\(m = n^\ell + \ell - 1\) and \(H_\ell (S) = 0\), so \((c H_\ell (S) + o
(\log n)) m = o (n^\ell \log n)\) for any $c$, but \(K (S) \in \Theta
(n^\ell \log n)\) in the expected case.

In this paper we investigate further the relationship between the order
$\ell$, the alphabet size $n$ and the string length $m$.  Our results
suggest $\ell$th-order empirical entropy stops being a reasonable
complexity metric for almost all strings about when $n^\ell$ surpasses
$m$.  For simplicity, we assume $\ell$ and $n$ are given to us as
(possibly constant) functions from $m$ to the positive integers and
consider \(S \in \{1, \ldots, n\}^m\).  In Section~\ref{upper_bounds} we
prove that, for any fixed \(c \geq 1\) and \(\epsilon > 0\), if \(n^{\ell
+ 1 / c} \log n \in o (m)\) and $m$ is sufficiently large, then \(K (S) <
(c H_\ell (S) + \epsilon) m\).  We use a new upper bound for compressing
probability distributions, which extends our results from~\cite{Gag06} and
may be of independent interest.  In Section~\ref{lower_bounds} we prove
that if \(\epsilon < 1 / c\), $\ell$ is fixed, \(n^{\ell + 1 / c -
\epsilon} \in \Omega (m)\) and $m$ is sufficiently large, then \(K (S) >
\left( c H_\ell (S) + \frac{\epsilon}{3} \log n \right) m\) with high
probability for randomly chosen $S$.  As a corollary we prove a nearly
matching lower bound for compressing probability distributions.

It seems interesting that slightly changing the relationship between
$\ell$, $n$ and $m$ can change \((c H_\ell (S) + o (\log n)) m\) from an
upper bound on \(K (S)\) to an almost certain lower bound.  Phenomena like
this one, in which small shifts in parameters change a property
asymptotically from very likely to very unlikely, are called
\emph{threshold phenomena}; they are common and well-studied in several
disciplines (see, e.g.,~\cite{KS06}) but we know of no others related to
data compression.  Although our proof of a threshold phenomenon requires
$\ell$ to be fixed, we emphasize it holds for any constant coefficient \(c
\geq 1\) before \(H_\ell (S)\) and any \(o (\log n)\) second term in the
formula.


\section{Upper bounds}
\label{upper_bounds}

We first rephrase the definition of empirical entropy: For \(\ell \geq
0\), the $\ell$th-order empirical entropy of a string $S$ is the minimum
self-information per character of $S$ emitted by an $\ell$th-order Markov
process.  The \emph{self-information} of an event with probability $p$ is
\(\log (1 / p)\).  An \emph{$\ell$th-order Markov process} is a string of
random variables in which each variable depends only on at most $\ell$
immediate predecessors (see, e.g.,~\cite{Sha48}); a process is said to
emit the values of its variables.  We use relative entropy~\cite{KL51},
also called the Kullback-Leibler distance, to prove the two definitions
equivalent.  Let \(P = p_1, \ldots, p_n\) and \(Q = q_1, \ldots, q_n\) be
probability distributions over \(\{1, \ldots, n\}\); the \emph{relative
entropy} between $P$ and $Q$, \(D (P \| Q) = \sum_{i = 1}^n p_i \log (p_i
/ q_i)\), is often used in information theory to measure how well $Q$
approximates $P$.  Although relative entropy is not a distance metric ---
it is not symmetric and does not obey the triangle inequality --- it is
$0$ when \(P = Q\) and positive otherwise.

\begin{thm}
\label{emp_ent}
For any string \(S \in \{1, \ldots, n\}^m\) and \(\ell \geq 0\), we have 
\(H_\ell (S) =\)
\linebreak
\(\frac{1}{m} \min \left\{ \rule{0ex}{2ex} \log (1 / \Pr [\mbox{\rm $Q$ 
emits $S$}])\,:\,\mbox{\rm $Q$ is an $\ell$th-order Markov process} 
\right\}\).
\end{thm}

\begin{pf}
Consider the probability an $\ell$th-order Markov process $Q$ emits $S$.  
Assume, without loss of generality, that $Q$ first emits \(s_1 \cdots
s_\ell\) with probability $1$.  For \(\alpha \in \{1, \ldots, n\}^\ell\),
let \(P_\alpha = p_{\alpha, 1}, \ldots, p_{\alpha, n}\) be the normalized
distribution of the characters in $S_\alpha$, so \(H (P_\alpha) = H_0
(S_\alpha)\); let \(Q_\alpha = q_{\alpha, 1}, \ldots, q_{\alpha, n}\),
where $q_{\alpha, a}$ is the probability $Q$ emits $a$ immediately after
an occurrence of $\alpha$.  Then
\begin{eqnarray*}
\lefteqn{\log \frac{1}{\Pr [\mbox{$Q$ emits $S$}]}}\\
& = & \log \prod_{i = \ell + 1}^m
	\frac{1}{q_{s_{i - \ell} \cdots s_{i - 1}, s_i}}\\
& = & \sum_{i = \ell + 1}^m
	\log \frac{1}{q_{s_{i - \ell} \cdots s_{i - 1}, s_i}}\\
& = & \sum_{|\alpha| = \ell} \sum_{a \in S_\alpha}
	\#_a (S_\alpha) \log \frac{1}{q_{\alpha, a}}\\
& = & \sum_{|\alpha| = \ell} |S_\alpha| \sum_{a \in S_\alpha}
	p_{\alpha, a} \left( \log \frac{p_{\alpha, a}}{q_{\alpha, a}} +
	\log \frac{1}{p_{\alpha, a}} \right)\\
& = & \sum_{|\alpha| = \ell} |S_\alpha|
	(D (P_\alpha \| Q_\alpha) + H (P_\alpha))\\
& \geq & \sum_{|\alpha| = \ell} |S_\alpha| H (P_\alpha)\\
& = & H_\ell (S) m\ ,
\end{eqnarray*}
with equality throughout if \(P_\alpha = Q_\alpha\) for \(\alpha \in \{1, 
\ldots, n\}^\ell\).
\qed
\end{pf}

We now consider how compactly we can store probability distributions,
Markov processes and, ultimately, strings.

\begin{lem}
\label{comp_pds}
Fix \(c \geq 1\) and \(\epsilon > 0\) and let \(P = p_1, \ldots, p_n\) be
a probability distribution over \(\{1, \ldots, n\}\).  For some
probability distribution $Q$ with \(D (P \| Q) < (c - 1) H (P)  +
\epsilon\), storing $Q$ takes \(O (n^{1 / c} \log n)\) bits.
\end{lem}

\begin{pf}
Let \(t \leq r n^{1 / c}\) be the number of probabilities in $P$ that are
at least \(\frac{1}{r n^{1 / c}}\), where \(r = \frac{2^{\epsilon / 
2}}{2^{\epsilon / 2} - 1}\).  For each such $p_i$, we record $i$ and 
\(\lfloor p_i r^2 n \rfloor\).  Since $r$ depends only on $\epsilon$, 
which is fixed, in total we use \(O (n^{1 / c} \log n)\) bits.  This 
information lets us later recover \(Q = q_1, \ldots, q_n\), where
\[q_i = \left\{ \begin{array}{ll}
\displaystyle \left( 1 - \frac{1}{r} \right) \frac{\lfloor p_i r^2 n \rfloor}
	{\sum \left\{ \lfloor p_j r^2 n \rfloor\,
	:\,p_j \geq \frac{1}{r n^{1 / c}} \right\}}
\hspace{3ex} & \mbox{if \(p_i \geq \frac{1}{r n^{1 / c}}\),}\\
\displaystyle \frac{1}{r (n - t)}
\hspace{3ex} & \mbox{otherwise.}
\end{array} \right. \]
Suppose \(p_i \geq \frac{1}{r n^{1 / c}}\); then \(p_i r^2 n \geq r\).  
Since \(\sum \left\{ \lfloor p_j r^2 n \rfloor\,:\,p_j \geq \frac{1} {r
n^{1 / c}} \right\} \leq r^2 n\),
\[p_i \log \frac{p_i}{q_i}
\leq p_i \log \left( \frac{r}{r - 1} \cdot
	\frac{p_i r^2 n}{\lfloor p_i r^2 n \rfloor} \right)
< 2 p_i \log \frac{r}{r - 1}
= p_i \epsilon\ .\]
Now suppose \(p_i < \frac{1}{r n^{1 / c}}\); then \(p_i \log (1 / p_i) > 
\frac{p_i}{c} \log n\).  Thus,
\[p_i \log \frac{p_i}{q_i}
< p_i \log \frac{r (n - t)}{r n^{1 / c}}
\leq \frac{(c - 1) p_i}{c} \log n
< (c - 1) p_i \log \frac{1}{p_i}\ .\]
Finally, since \(p \log (1 / p) \geq 0\) for \(p \leq 1\), we have
\[D (P \| Q)
< \sum \left\{ (c - 1) p_i \log \frac{1}{p_i}\,
	:\,p_i < \frac{1}{r n^{1 / c}} \right\} + \epsilon
\leq (c - 1) H (P) + \epsilon\ .\]
\qed
\end{pf}

\begin{cor}
\label{comp_mps}
Fix \(c \geq 1\) and \(\epsilon > 0\) and consider a string \(S \in \{1,
\ldots, n\}^m\).  For some $\ell$th-order Markov process $Q$ with \(\log
(1 / \Pr [\mbox{\rm $Q$ emits $S$}]) < (c H_\ell (S) + \epsilon) m\), 
storing $Q$ takes \(O (n^{\ell + 1 / c} \log n)\) bits.
\end{cor}

\begin{pf}
First we store \(s_1 \cdots s_\ell\).  For \(\alpha \in \{1, \ldots,
n\}^\ell\), let \(P_\alpha = p_{\alpha, 1}, \ldots, p_{\alpha, n}\) be the
normalized distribution of characters in $S_\alpha$ and let \(Q_\alpha =
q_{\alpha, 1}, \ldots, q_{\alpha, n}\) be the probability distribution
with \(D (P_\alpha \| Q_\alpha) < (c - 1) H (P_\alpha) + \epsilon\)
obtained from applying Lemma~\ref{comp_pds} to $c$, $\epsilon$ and
$P_\alpha$.  We store every $Q_\alpha$, using a total of \(O (n^{\ell + 1 
/ c} \log n)\) bits.

This information lets us later recover a Markov process $Q$ that first
emits \(s_1 \cdots s_\ell\) and in which, for \(\alpha \in \{1, \ldots,
n\}^\ell\) and \(a \in \{1, \ldots, n\}\), the probability $a$ is emitted
immediately after an occurrence of $\alpha$ is $q_{\alpha, a}$.  As in the
proof of Theorem~\ref{emp_ent}, \(\log (1 / \Pr [\mbox{$Q$ emits $S$}]) =
\sum_{|\alpha| = \ell} |S_\alpha| (D (P_\alpha \| Q_\alpha) + H
(P_\alpha))\), so
\[\log \frac{1}{\Pr [\mbox{$Q$ emits $S$}]}
< \sum_{|\alpha| = \ell} |S_\alpha| (c H (P_\alpha) + \epsilon)
\leq (c H_\ell (S) + \epsilon) m\ .\]
\qed
\end{pf}

We note that, given a string \(S \in \{1, \ldots, n\}^m\), we can store an
$\ell$th-order Markov process $Q$ with \(\log (1 / \Pr [\mbox{$Q$ emits
$S$}]) = H_\ell (S)\) in \(O \left( n^{\ell + 1} \log \left(
\frac{m}{n^{\ell + 1}} + 1 \right) \right)\) bits, as a table containing
\(\#_a (S_\alpha) = \#_{\alpha a} (S) \leq m\) for \(\alpha a \in \{1,
\ldots, n\}^{\ell + 1}\).  Grossi, Gupta and Vitter~\cite{GGV??}
investigated the space needed for such a table; they also showed that,
apart from the cost of storing the table, we can store $S$ in \(H_\ell
(S)\) bits.  However, because we do not see how to store the table in less
space when there is a constant coefficient \(c > 1\) before \(H_\ell
(S)\), we tolerate the $\epsilon$ term in Corollary~\ref{comp_mps} and the
following theorem.

\begin{thm}
\label{comp_strings}
Fix \(c \geq 1\) and \(\epsilon > 0\) and let $\ell$ and $n$ be functions
from $m$ to the positive integers.  Consider a string \(S \in \{1, \ldots,
n\}^m\).  If \(n^{\ell + 1 / c} \log n \in o (m)\) and $m$ is sufficiently 
large, then \(K (S) < (c H_\ell (S) + \epsilon) m\).
\end{thm}

\begin{pf}
By Corollary~\ref{comp_mps}, since \(n^{\ell + 1 / c} \log n \in o (m)\)  
and $m$ is sufficiently large, we can store an $\ell$th-order Markov
process $Q$ with \(\log (1 / \Pr [\mbox{$Q$ emits $S$}]) < (c H_\ell (S) +
\epsilon / 2) m\) in \(\epsilon m / 2 - 1\) bits.  Shannon~\cite{Sha48}
showed how, given $Q$, we can store $S$ in \(\left\lceil \log (1 / 
\Pr[\mbox{$Q$ emits $S$}]) \right\rceil\) bits.  Thus, we can store $Q$ 
and $S$ together in fewer than \((c H_\ell (S) + \epsilon) m\) bits.
\qed
\end{pf}


\section{Lower bounds}
\label{lower_bounds}

Consider the so-called \emph{birthday paradox}: If we draw $m$ times from
\(\{1, \ldots, n\}\), then the probability at least two of the numbers
drawn are the same is about \(1 - 1 / e^{\frac{m (m - 1)}{2 n}}\).  Thus,
for \(\ell \geq 1\), if \(n^{1 / 2} \in \omega (m)\) and $S$ is chosen
randomly, then with high probability \(H_\ell (S) = 0\) because no
character appears more than once in $S$.  (Notice also \(H_0 (S) \leq \log
m \leq \log (n) / 2\)  for sufficiently large $m$.)  Thus, we cannot lift
the restriction on $n$ and $\ell$ in Theorem~\ref{comp_strings} to \(n^{1
/ 2 - \epsilon} \in O (m)\).  We use a similar but more complicated
argument to show we cannot even lift the restriction to \(n^{\ell + 1 / c
- \epsilon} \in O (m)\).  Essentially, we use a Chernoff bound on the
probability of there being any frequent $\ell$-tuples in $S$.  Since the
probability of an $\ell$-tuple occurring somewhere in $S$ depends on
whether it occurs in neighbouring positions, we apply the following
intuitive lemma (proven in, e.g.,~\cite{MR95}) before we apply the
Chernoff bound.

\begin{lem}
\label{x_y_lemma}
Let \(X_1, \ldots, X_m\) be binary random variables such that, for \(1 
\leq i \leq m\) and \(b \in \{0, 1\}^{i - 1}\), \(\Pr \left[ X_i = 
1\,\left|\,\rule{0ex}{2ex} \right. X_1 \cdots X_{i - 1} = b \right] \leq 
p\).  Let \(Y_1, \ldots, Y_m\) be independent binary random variables, 
each equal to $1$ with probability $p$.  For \(0 \leq q \leq 1\),
\[\Pr \left[ \sum_{j = 1}^m X_j \geq q m \right]
\leq \Pr \left[ \sum_{j = 1}^m Y_j \geq q m \right]\ .\]
\end{lem}

\begin{thm}
\label{incomp_strings}
Fix \(c \geq 1\), $\epsilon$ with \(0 < \epsilon < 1 / c\) and \(\ell
\geq 1\) and let $n$ be a function from $m$ to the positive integers.  
Choose a string \(S \in \{1, \ldots, n\}^m\) uniformly at random.  If
\(n^{\ell + 1 / c - \epsilon} \in \Omega (m)\) and $m$ is sufficiently
large, then \(K (S) > \left( c H_\ell (S) + \frac{\epsilon}{3} \log n
\right) m\) with high probability.
\end{thm}

\begin{pf}
Since there are $n^m$ choices for $S$ and only
\[\sum \left\{ 2^i\,:\,
	0 \leq i \leq \lfloor (1 - \epsilon / 3) m \log n \rfloor \right\}
\leq 2 n^{(1 - \epsilon / 3) m} - 1\]
binary strings of length at most \((1 - \epsilon / 3) m \log n\), we have
\(K (S) \geq (1 - \epsilon / 3)\)
\linebreak
\(m \log n\) with probability greater than \(1 - 2 / n^{\epsilon m / 3}\).  
Thus, we need only show \(c H_\ell (S) < (1 - 2 \epsilon / 3) \log n\)
with high probability.  By definition,
\begin{eqnarray*}
\lefteqn{H_\ell (S)}\\
& \leq & \max_{|\alpha| = \ell} \{H_0 (S_\alpha)\}\\
& \leq & \max_{|\alpha| = \ell} \{\log |\{a\,:\,a \in S_\alpha\}|\}\\
& \leq & \max_{|\alpha| = \ell} \left\{ \log \left( \rule{0ex}{2ex}
	|\{a\,:\,a \in S_\alpha, a \not \in \alpha\}| +
	\ell \right) \right\}\ .
\end{eqnarray*}
Notice \(n \in \omega \left( m^{\frac{1}{\ell + 1 / c}} \right)\).  We 
will show
\[\Pr \left[ |\{ a\,:\,a \in S_\alpha, a \not \in \alpha\}|
	\geq n^{1 / c - 2 \epsilon / 3} - \ell \right]
\leq \frac{1}{2^{n^{\epsilon / 3} - \ell}}\]
for each \(\alpha \in \{1, \ldots, n\}^\ell\), so
\begin{eqnarray*}
\Pr \left[ \max_{|\alpha| = \ell}
	\{|\{ a\,:\,a \in S_\alpha, a \not \in \alpha\}|\}
	\geq n^{1 / c - 2 \epsilon / 3} - \ell \right]
& \leq & \frac{n^\ell}{2^{n^{\epsilon / 3} - \ell}}\ ,\\
&& \\
\Pr \left[ \max_{|\alpha| = \ell} \left\{ \log \left( \rule{0ex}{2ex}
	|\{ a\,:\,a \in S_\alpha, a \not \in \alpha\}| + \ell \right) \right\}
	\geq \left( \frac{1}{c} - \frac{2 \epsilon}{3} \right)
	\log n \right]
& \leq & \frac{n^\ell}{2^{n^{\epsilon / 3} - \ell}}
\end{eqnarray*}
and \(c H_\ell (S) < (1 - 2 \epsilon c / 3) \log n \leq (1 - 2 \epsilon 
/ 3) \log n\) with high probability.

Consider \(\alpha \in \{1, \ldots, n\}^\ell\).  Let \(X_1, \ldots, X_{m -
\ell}\) be binary random variables, with \(X_i = 1\) if \(s_i \cdots s_{i
+ \ell - 1} = \alpha\) and \(s_{i + \ell} \not \in \alpha\).  Notice
\(|\{a\,:\,a \in S_\alpha, a \not \in \alpha\}| + \ell \leq \sum_{i =
1}^{m - \ell} X_i + \ell\).  For \(\ell + 1 \leq i \leq m - \ell\), by
definition, $X_i$ is independent of \(X_1, \ldots, X_{i - \ell - 1}\); if
any of \(X_{i - \ell}, \ldots, X_{i - 1}\) are 1, then at least one of
\(s_i, \ldots, s_{i + \ell - 1}\)  is not in $\alpha$, so \(X_i = 0\); and
\begin{eqnarray*}
\lefteqn{\Pr \left[ X_i = 1\,\left|\,\rule{0ex}{2ex} \right.
	X_{i - \ell} = \cdots = X_{i - 1} = 0 \right]}\\
& = & \frac{\Pr \left[ X_i = 1\,\mbox{and}\,
	X_{i - \ell} = \cdots = X_{i - 1} = 0 \right]}
	{\Pr \left[ X_{i - \ell} = \cdots = X_{i - 1} = 0 \right]}\\
& = & \frac{\Pr [X_i = 1]}
	{1 - \Pr \left[
	X_{i - \ell} = 1\,\mbox{or \dots\,or}\,X_{i - 1} = 1 \right]}\\
& \leq & \frac{\Pr [X_i = 1]}
	{1 - \sum_{j = i - \ell}^{i - 1} \Pr [X_j = 1]}\\
& \leq & \frac{1 / n^\ell}{1 - \ell / n^\ell}\\
& = & \frac{1}{n^\ell - \ell}\ .
\end{eqnarray*}
Let \(Y_1, \ldots, Y_{m - \ell}\) be independent binary random variables,
each equal to $1$ with probability \(p = \frac{1}{n^\ell - \ell}\), and
let \(q = \frac{n^{1 / c - 2 \epsilon / 3} - \ell}{m - \ell}\).  If \(q >
1\) the proof is finished, because \(\Pr \left[ \sum_{i = 1}^{m - \ell} 
X_i \geq q (m - \ell) \right] = 0\); otherwise by 
Lemma~\ref{x_y_lemma},
\[\Pr \left[ \sum_{i = 1}^{m - \ell} X_i
	\geq q (m - \ell) \right]
\leq \Pr \left[ \sum_{i = 1}^{m - \ell} Y_i
	\geq q (m - \ell) \right]\]
and it remains for us to show
\[\Pr \left[ \sum_{i = 1}^{m - \ell} Y_i
	\geq q (m - \ell) \right]
\leq \frac{1}{2^{n^{\epsilon / 3} - \ell}}\ .\]
Since $\ell$ is fixed and \(n^{\ell + 1 / c - \epsilon} \in \Omega (m)\), 
we have \(p (m - \ell) \in O (n^{1 / c - \epsilon}) \subset\)
\linebreak 
\(o (q (m - \ell))\); thus, for sufficiently large $m$, \(q (m - \ell) 
\geq 6 p (m - \ell)\) and we can use the following simple Chernoff 
bound~\cite{HR90}:
\[\Pr \left[ \sum_{i = 1}^{m - \ell} Y_i \geq q (m - \ell) \right]
\leq \frac{1}{2^{q (m - \ell)}}
= \frac{1}{2^{n^{1 / c - 2 \epsilon / 3} - \ell}}\ .\]
Finally, since \(\epsilon < 1 / c\),
\[\Pr \left[ \sum_{i = 1}^{m - \ell} Y_i \geq q (m - \ell) \right]
\leq \frac{1}{2^{n^{\epsilon / 3} - \ell}}\ .\]
\qed
\end{pf}

\pagebreak

\begin{cor}
\label{incomp_pds}
Fix \(c \geq 1\) and $\epsilon$ with \(0 < \epsilon < 1 / c\) and let $P$
be a probability distribution over \(\{1, \ldots, n\}\).  In the worst
case, for any probability distribution $Q$ with \(D (P \| Q) \leq (c - 1)  
H (P) + o (\log n)\), storing $Q$ takes \(\omega (n^{1 / c - \epsilon})\)
bits.
\end{cor}

\begin{pf}
For the sake of a contradiction, assume there exists an algorithm
$\mathcal{A}$ that, given any probability distribution $P$ over \(\{1,
\ldots, n\}\), stores a probability distribution $Q$ with \(D (P \| Q)
\leq (c - 1) H (P) + o (\log n)\) in \(O (n^{1 / c - \epsilon})\) bits.  
Then a proof similar to that of Theorem~\ref{comp_strings}, but
substituting $\mathcal{A}$ for Lemma~\ref{comp_pds}, yields:
\begin{quotation}
\noindent \it Fix \(c \geq 1\) and $\epsilon$ with \(0 < \epsilon < 1 /
c\) and let $\ell$ and $n$ be functions from $m$ to the positive integers.  
Consider a string \(S \in \{1, \ldots, n\}^m\).  If \(n^{\ell + 1 / c -
\epsilon} \in o (m)\), then \(K (S) \leq (c H_\ell (S) + o (\log n)) m\).
\end{quotation}
Suppose we fix $c$ and $\ell$, choose \(\epsilon < 1 / c\) and $n$ such
that \(n^{\ell + 1 / c - \epsilon} \in o (m)\) but \(n^{\ell + 1 / c -
\epsilon / 2} \in \Omega (m)\), and choose a string \(S \in \{1, \ldots,
n\}^m\)  uniformly at random.  The claim above gives \(K (S) \leq (c
H_\ell (S) + o (\log n)) m\) but by Theorem~\ref{incomp_strings}, for
sufficiently large $m$, \(K (S) > \left( c H_\ell (S) + \frac{\epsilon}{6}
\log n \right) m\) with high probability.
\qed
\end{pf}


\section{Future work}
\label{future_work}

Suppose we want to store a probability distribution $P$ over a set of 
strings.  We recently proved that, in theory, if the relative entropy is 
small between $P$ and the probability distribution induced by a low-order 
Markov process $Q$, the we can store $P$ accurately and efficiently by 
storing an approximation of $Q$.  We hope experiments will show this 
technique to be practical.

Our proof of Theorem~\ref{incomp_strings} is slightly complicated because 
if, for some $\ell$-tuple $\alpha$, a non-empty string is both a suffix 
and a prefix of $\alpha$, then occurrences of $\alpha$ can overlap and any 
one occurrence increases the probability of others.  In this paper we used 
the fact that if two occurrences of $\alpha$ overlap, the the first must 
be immediately followed by a character in $\alpha$.  We recently proved 
that, moreover, it must be immediately followed by one of \(O (\log 
\ell)\) characters.  We are now trying to use this result to prove a 
version of Theorem~\ref{incomp_strings} that does not require $\ell$ to be 
fixed.

We are also trying another approach to generalize
Theorem~\ref{incomp_strings}.  Results about linear de Bruijn sequences
are often proved by considering them as Eulerian tours on certain graphs,
called de Bruijn graphs.  In fact, any string can be considered as a walk
on a de Bruijn graph; random strings correspond to random walks.  Since de
Bruijn graphs are good expanders, random walks on them have properties
that may be useful in reasoning about random strings.


\section*{Acknowledgments}
\label{acknowledgments}

Many thanks to Giovanni Manzini and Charlie Rackoff, who supervised this
research; Mark Braverman, Paolo Ferragina, Roberto Grossi and the
anonymous reviewers, for helpful comments; and Alistair Moffat, for
editorial patience.


\bibliographystyle{plain}
\bibliography{large}

\end{document}